\documentclass[12pt,preprint]{aastex}

\begin{document}

\title{Detection of a tertiary brown dwarf companion in the sdB-type eclipsing binary HS\,0705+6700}

\author{Qian S.-B.\altaffilmark{1,2,3}, Zhu
L.-Y.\altaffilmark{1,2,3}, Zola, S.\altaffilmark{4,5}, Liao
W.-P.\altaffilmark{1,2,3}, Liu L.\altaffilmark{1,2,3}, Li
L.-J.\altaffilmark{1,2,3}, Winiarski M.\altaffilmark{5}, Kuligowska
E.\altaffilmark{4}, and Kreiner J. M.\altaffilmark{5}}

\altaffiltext{1}{National Astronomical Observatories/Yunnan
Observatory, Chinese Academy of Sciences (CAS), P.O. Box 110, 650011
Kunming, P.R. China (e-mail: qsb@ynao.ac.cn)}

\altaffiltext{2}{United Laboratory of Optical Astronomy, Chinese
Academy of Sciences (ULOAC), 100012 Beijing, P. R. China}

\altaffiltext{3}{Graduate University of the Chinese Academy of
Sciences, 10049 Beijing, P. R. China}

\altaffiltext{4}{Astronomical Observatory, Jagiellonian University,
ul. Orla 171, 30-244 Krakow, Poland}

\altaffiltext{5}{Mt. Suhora Observatory, Cracow Pedagogical
University ul. Podchorazych 2, 30-084 Krakow, Poland}

\begin{abstract}
HS\,0705+6700 is a short-period (P=2.3\,hours), close binary containing a hot
sdB-type primary and a fully convective secondary.
 We have monitored this eclipsing binary for more than 2 years
and as a result, 32 times of light minimum were obtained. Based on
our new eclipse times together with these compiled from the
literature, it is discovered that the O-C curve of HS\,0705+6700
shows a cyclic variation with a period of 7.15\,years and a
semiamplitude of 92.4\,s. The periodic change was analyzed for the
light-travel time effect that may be due to the presence of a
tertiary companion. The mass of the third body is determined to be
$M_3\sin{i^{\prime}}=0.0377(\pm0.0043)$\,$M_{\odot}$ when a total
mass of $0.617$\,$M_{\odot}$ for HS\,0705+6700 is adopted. For
orbital inclinations $i^{\prime}\ge{32.8}^{\circ}$, the mass of the
tertiary component would be below the stable hydrogen-burning limit
of $M_3\sim0.072$\,$M_{\odot}$, and thus it would be a brown dwarf.
The third body is orbiting the sdB-type binary at a distance shorter
than 3.6 astronomical units (AU). HS\,0705+6700 was formed through
the evolution of a common envelope after the primary becomes a red
giant. The detection of a sub-stellar companion in HS\,0705+6700
system at this distance from the binary could give some constraints
on stellar evolution in such systems and the interactions between
red giants and their companions.
\end{abstract}


\section{Introduction}

HS\,0705+6700 (=GSC\,4123-265) was listed as a dwarf candidate from
the Hamburg Schmidt survey (Hagen et al. 1995). Follow-up
spectroscopy by Heber et al. (1999) and Edelmann et al. (2001)
revealed that its effective temperature lies in the predicted
pulsational instability. Therefore, in order to search for
pulsations, this star was included in a photometric monitoring
programme at the Nordic Optical Telescope (see Ostensen et al.
2001a, b). The observations indicated that it was an eclipsing
binary (Drechsel et al. 2001). A detailed photometric and
spectroscopic investigation was carried out by Drechsel et al.
(2001) who discovered that HS\,0705+6700 is a detached, short-period
eclipsing binary. Absolute parameters of both components were
determined suggesting the primary is a subluminous B (sdB) star,
while the secondary is a cool stellar object that does not
contribute to the total optical light apart from a strong reflection
effect. These detections reveal that HS\,0705+6700 is the third one
of a small group of HW Vir-like eclipsing binary stars that consists
of a very hot sdB type primary component and a fully convective
M-type secondary with a period between 2 and 3 hours. Up to now,
only six of this type of binaries have been discovered (e.g.,
Menzies \& Marang 1986; Kilkenny et al. 1998; Drechsel et al. 2001;
Ostensen et al. 2007; Polubek et al. 2007; Wils et al. 2007). The
hot sdB components in this group of binaries are on the extreme
horizontal branch of the Hertzsprung-Russell diagram, they burn
helium in their cores, and have very thin hydrogen envelopes. They
are formed through a common-envelope evolution (e.g., Han et al.
2003) and will evolve into normal cataclysmic variables (CV) (e.g.,
Shimansky et al. 2006).

As pointed out by Qian et al. (2008a), because of the compact
structures and large temperature differences between the components,
light curves of this group of binaries show a strong reflection
effect with very sharp primary and shallow secondary minima.
Therefore, eclipse times can be determined with a high precision
(e.g., Kilkenny et al. 1994, 2000), and very small-amplitude orbital
period variations could be detected by analyzing the
observed-calculated (O-C) diagram. Orbital period variations of HW
Vir, the prototype of this group of systems, were discovered (e.g.,
Kilkenny et al. 1994; Qian et al. 2008a; Lee et al. 2009), which
show a combination of a cyclic variation and a long-term period
decrease. The cyclic variation suggests the presence of a brown
dwarf tertiary companion in the system, while the continuous
decrease can be explained as secular angular momentum loss via
magnetic braking of the fully convective component star or as a part
of another long-period cyclic change via the existence of another
companion. To search for the variations in the orbital period of
HS\,0705+6700, it has been monitored since 2006. Here we report the
discovery of a cyclic change in the orbital period of HS\,0705+6700
that reveals the presence of a tertiary, most likely a brown dwarf
companion in this system.

\section{New observations and the orbital period change of HS\,0705+6700}

\begin{table}
\caption{New CCD times of light minimum for HS\,0705+6700.}
\begin{center}
\begin{small}
\begin{tabular}{llllll}\hline\hline
J.D. (Hel.)  & Errors & Min. & Filters & E & Telescopes\\
+2400000 (days) & days &  & & & \\ \hline
54081.59917 &  $\pm0.00005 $ & II & R  &23616.5& Suhora60 \\
54081.64687 &  $\pm0.00018 $ & I  & R  &23617  & Suhora60  \\
54420.23535 &  $\pm0.00015 $ & I  & VR & 27157 &Xinglong85 \\
54420.28360 &  $\pm0.00025 $ & II & VR & 27157.5&Xinglong85 \\
54492.06611 &  $\pm0.00018 $ & I  & RI & 27908 &Xinglong85 \\
54492.11397 &  $\pm0.00022 $ & II & RI & 27908.5&Xinglong85 \\
54492.16176 &  $\pm0.00018 $ & I  & RI & 27909 & Xinglong85 \\
54517.41211 &  $\pm0.00034 $ & I  & BG40&28173 & Suhora60 \\
54517.50821 &  $\pm0.00037 $ & I  & BG40&28174 & Suhora60 \\
54642.42239 &  $\pm0.00027 $ & I  & BG40&29480 & Suhora60  \\
54659.44794 &  $\pm0.00047 $ & I  & BG40&29658 & Suhora60 \\
54684.41068 &  $\pm0.00024 $ & I  & BG40&29919 & Krakow50 \\
54684.50649 &  $\pm0.00024 $ & I  & None&29920 & Krakow50 \\
54706.40500 &  $\pm0.00029 $ & I  & BG40&30149 & Krakow50 \\
54706.50400 &  $\pm0.00036 $ & I  & BG40&30150 & Krakow50 \\
54715.49683 &  $\pm0.00020 $ & I  & BG40&30244 & Suhora60 \\
54715.59165 &  $\pm0.00023 $ & I  & BG40&30245 & Suhora60 \\
54718.55611 &  $\pm0.00041 $ & I  & BG40&30276 & Suhora60 \\
54729.31741 &  $\pm0.00015 $ & II & V  &30388.5& Xinglong85 \\
54729.36497 &  $\pm0.00015 $ & I  & V  &30389 & Xinglong85\\
54741.32088 &  $\pm0.00008 $ & I  & V  &30514  &Xinglong85 \\
54741.36856 &  $\pm0.00011 $ & II & V  &30514.5&Xinglong85 \\
54745.33801 &  $\pm0.00020 $ & I  & BG40&30556 & Krakow50  \\
54760.64233 &  $\pm0.00032 $ & I  & BG40&30716 & Suhora60   \\
54761.40648 &  $\pm0.00023 $ & I  & None&30724 & Krakow50     \\
54780.24945 &  $\pm0.00015 $ & I  & BG40&30921 & Krakow50   \\
54780.24902 &  $\pm0.00009 $ & I  & V  & 30921 & Xinglong85 \\
54780.34421 &  $\pm0.00034 $ & I  & BG40&30922 & Krakow50   \\
54808.27334 &  $\pm0.00034 $ & I  & V  & 31214 & Xinglong60 \\
54808.36919 &  $\pm0.00031 $ & I  & V  & 31215 & Xinglong60 \\
54810.47380 &  $\pm0.00020 $ & I  & Luminance & 31237 & Krakow50 \\
54810.56997 &  $\pm0.00030 $ & I  & R  & 31238 & Xuhora60  \\
54815.25580 &  $\pm0.00010 $ & I  & V  & 31287 & Xinglong85 \\
54815.35156 &  $\pm0.00014 $ & I  & V  & 31288 & Xinglong85 \\
54817.26439 &  $\pm0.00039 $ & I  & V  & 31308 & Xinglong60 \\
54817.36029 &  $\pm0.00039 $ & I  & V  & 31309 & Xinglong60 \\
54829.41098 &  $\pm0.00019 $ & I  & BG40&31435 & Xuhora60 \\
54838.21093 &  $\pm0.00012 $ & I  & V  & 31527 & Xinglong85 \\
\hline
\end{tabular}
\tablecomments{Xinglong85 and Xinglong60: the 85-cm and the 60-cm
telescopes in Xinglong station of National Astronomical
Observatories (NAO), Suhora60: the 60-cm Mt. Suhora telescope,
Krakow50: the 50-cm Krakow telescope.}
\end{small}
\end{center}
\end{table}

\begin{figure}
\begin{center}
\includegraphics[angle=0,scale=1.0]{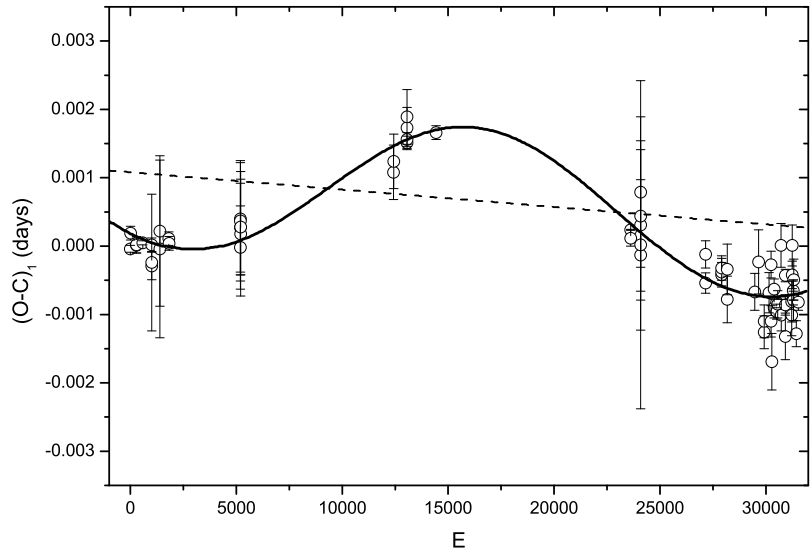}
\caption{A plot of the $(O-C)_1$ diagram of HS\,0705+6700 with
respect to the linear ephemeris given by Drechsel et al. (2001). The
solid line suggests a combination of a revised linear ephemeris and
a cyclic change, while the dashed line refers to the revision of the
orbital period. }
\end{center}
\end{figure}

\begin{figure}
\begin{center}
\includegraphics[angle=0,scale=1.0]{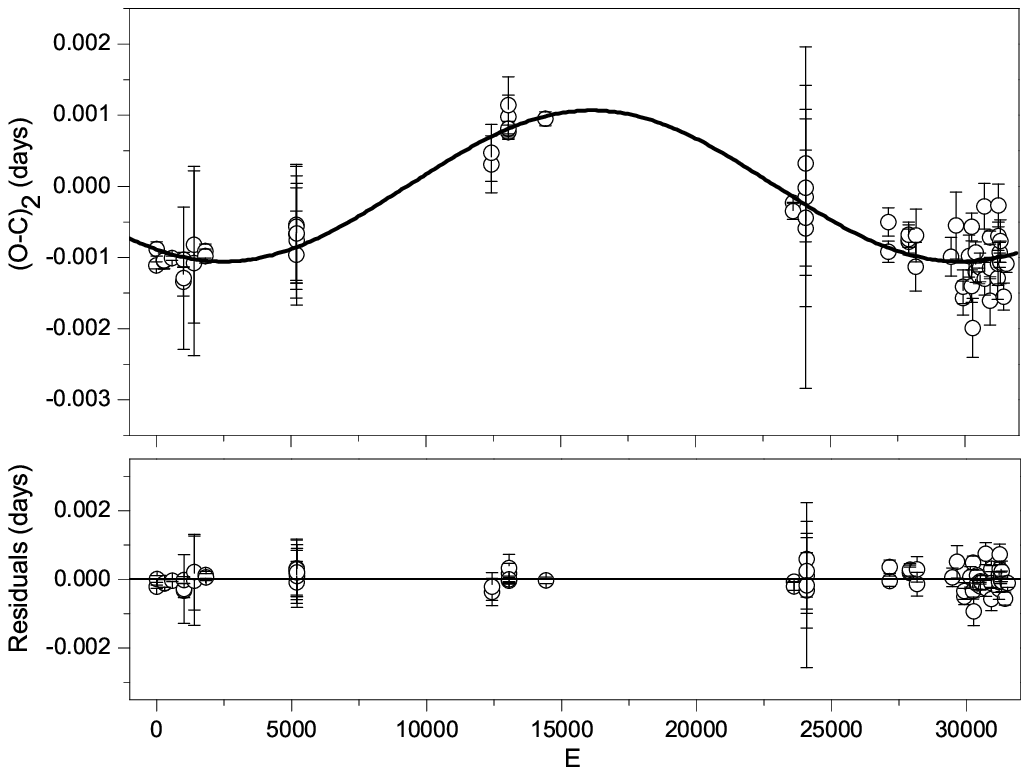}
\caption{The $(O-C)_2$ curve of HS\,0705+6700 with respect to the
new linear ephemeris in Eq.(2) is shown in the upper panel where the
periodic variation can be seen more clearly. After the
small-amplitude period oscillation was removed, the residuals are
shown in the lower panel where no changes can be traced.}
\end{center}
\end{figure}

Drechsel et al. (2001) published 13 times of light minimum of
HS\,0705+6700 and obtained the first linear ephemeris,
\begin{equation}
Min. I = HJD\,2451822.75982+0.09564665\times{E},
\end{equation}
where HJD\,2451822.75982 is the initial epoch and 0.09564665 is the
orbital period. Later, some eclipse times were derived by Niarchos
et al. (2003), N\'{e}meth et al. (2005), and Kruspe et al. (2007).
To search for the variations in the orbital period of HS\,0705+6700,
it was monitored from December, 2006 to December, 2008 by using four
telescopes in China and Poland (the 85-cm and the 60-cm telescopes
in Xinglong station of National Astronomical Observatories (NAO),
and the 60-cm Mt. Suhora and the 50-cm Krakow telescopes in Poland).
38 eclipse times were obtained and they are listed in Table 1. The
$(O-C)_1$ values of all available times of  minima were calculated
by using the ephemeris from Eq. (1). The corresponding $(O-C)_1$
diagram is shown in Fig. 1, where our 38 new minima times and the
other 31 eclipse times collected from the sources noted above.

As shown in Figure 1, the linear component of the orbital period of
HS\,0705+6700 needs revision and it appears that there is a cyclic
variation as well. To describe the general $(O-C)_1$ trend
satisfactorily, a new linear ephemeris is required (dashed line in
Fig. 1) with additional cyclic variations superimposed. Using the
least squares method, we determined,
\begin{eqnarray}
Min. I&=&2451822.76090(\pm0.00007)\nonumber\\
     && +0.095646625(\pm0.000000003)\times{E}\nonumber\\
    & &+0.00107(\pm0.00007)\sin[0.^{\circ}0132(\pm0.0001)\times{E}+237.^{\circ}2(\pm3.^{\circ}6)].
\end{eqnarray}
The derived orbital period is slightly shorter than that determined
by Drechsel et al. (2001). The cyclic oscillation has an amplitude
of 92.4 seconds and a period of 7.15 years. During the analysis, two
timings of light minima, HJD\,2451957.5274 and HJD\,2454706.50400,
were not used because their $(O-C)_1$ values show large scatter when
compared with the general trend formed by the other data points.
Actually, the eclipse minimum, HJD\,2451957.5274, has been deleted
by Drechsel et al. (2001) too, in their analysis.

The $(O-C)_2$ values calculated with the new linear ephemeris are
plotted in the upper panel of Figure 2 where the cyclic change is
seen more clearly. After the periodic change was subtracted from the
$(O-C)_2$ curve, the residuals are displayed in the lower panel
where no variations can be found indicating that equation (2) gives
a good fit to the $(O-C)_1$ curve.

\section{Discussions and conclusions}

\begin{figure}
\begin{center}
\includegraphics[angle=0,scale=1.0]{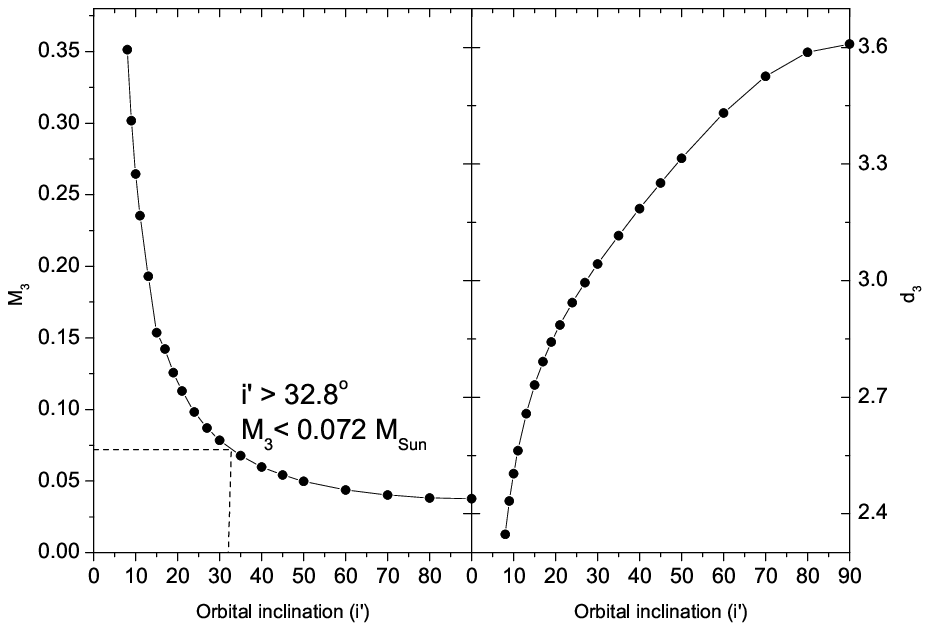}
\caption{The relations between the mass $M_3$\,($M_{\odot}$) and the
orbital radius $d_3$ (AU) of the tertiary component and its orbital
inclination $i^{\prime}$ in the HS\,0705+6700 system. The tertiary
companion should be a brown dwarf when the orbital inclination is
larger than $32.8^{\circ}$, while the orbital radius $d_3$ of the
tertiary component is always less than 3.6\,AU. }
\end{center}
\end{figure}

One cause of cyclic period change could be the magnetic activity
cycles of the fully convective component (i. e., the Applegate
mechanism) (Applegate 1992). It is assumed in the mechanism that a
certain amount of angular momentum is periodically exchanged between
the inner and the outer parts of the convection zone, and therefore
the rotational oblateness and thus the orbital period will vary when
the cool component goes through its activity cycles. As in the cases
of HW Vir and NN Ser (Qian et al. 2008a; Brinkworth et al. 2005),
the fully convective secondary in HS\,0705+6700 rotates mainly as a
rigid body, and lacks the thin interface layer between a radiative
core and a convective envelope, where dynamo processes are thought
to concentrate at for solar-type stars (e.g., Barnes 2005). The
analyses for HW Vir and NN Ser indicated that the required energies
are much larger than the total radiant energy of the M-type
components, suggesting, that the mechanism of Applegate can not
interpret the cyclic period variations of the two systems. Moreover,
as discussed by Qian et al. (2008a, b), a more general explanation
of the cyclic period changes in close binaries would be the
light-travel time effect via the presence of a third body.

Therefore, we analyzed HS\,0705+6700 for the light-time effect that
arises from the gravitational influence of a third companion. The
presence of a tertiary body produces the relative distance changes
of the eclipsing pair as it orbits the barycenter of the triple
system. Since the sine fit seems quite good, we assumed the orbit of
the third body to be circular. With the absolute parameters
determined by Drechsel et al. (2001), we derived the mass function
and the mass of the tertiary companion as:
$f(m)=1.25(\pm0.24)\times{10^{-5}}\,M_{\odot}$ and
$M_3\sin{i}^{\prime}=0.0377(\pm0.0043)\,M_{\odot}$, respectively.
The relations between the mass $M_3$ and the orbital radius $d_3$ of
the tertiary component and its orbital inclination $i^{\prime}$ are
displayed in Figure 3. When the orbital inclination of the third
body is larger than $32.8^{\circ}$, the mass of the tertiary
component corresponds to
$0.0377\,M_{\odot}\le{M_3}\le0.072\,M_{\odot}$. In this case, the
tertiary component can not undergo a stable hydrogen burning in the
core, and it should be a brown dwarf. Therefore, with 63.6\%
probability, the third body is a substellar object (by assuming a
random distribution of orbital plane inclination). However,
depending on the unknown orbital inclination of the third body, a
low-mass, stellar companion cannot be totally excluded but with a
lower possibility of 36.4\%.

HS\,0705+6700 has passed through the phase of a common envelope (CE)
after the more massive component star in the original system evolves
into a red giant. The ejection of CE removed a large amount of the
angular momentum, and the present, short-period sdB-type binary has
been formed. As it is shown in Figure 3, the orbital radius $d_3$ of
the tertiary component is smaller than 3.6\,AU. The detection of a
brown dwarf or a very low-mass stellar companion in HS\,0705+6700 at
this distance, could give some constraints on the stellar evolution
and the interaction between red giants and their companions.

Apart from cyclic period changes, a long-term period decrease was
discovered in HW Vir that can be plausibly explained by secular
angular momentum loss via magnetic braking of its fully convective
component (Qian et al. 2008a; Lee et al. 2009). If this is true, a
long-term period decrease could be discovered in HS\,0705+6700.
Actually, as displayed in Figure 1, the $(O-C)_1$ diagram can also
be described by a combination of a cyclic change and a long-term
period decrease. To check wether a long-term period decrease exists
or not, more times of light minimum are required in the future.

\acknowledgments{This work is partly supported by Chinese Natural
Science Foundation (No.10778718 and No.10778707), the National Key
Fundamental Research Project through grant 2007CB815406, the Yunnan
Natural Science Foundation (No. 2008CD157), and by the Special
Foundation of President and West Light Foundation of the Chinese
Academy of Sciences. New CCD photometric observations of the system
were obtained with several small telescopes in China and in Poland.
The authors thank the referee for those useful comments and
suggestions that help to improve the original manuscript. }

\end{document}